\documentclass[preprint,5p,times,twocolumn,longtitle]{elsarticle}

\usepackage{amsmath}
\usepackage{amssymb}
\usepackage{hyperref}

\journal{Physics Letters B}

\begin{document}

\begin{frontmatter}

\title{Search for $C$ violation in the decay $\eta\rightarrow\pi^0+e^++e^-$
with WASA-at-COSY}

\author[IKPUU]{The WASA-at-COSY Collaboration\\[2ex] P.~Adlarson}
\author[ASWarsN]{W.~Augustyniak}
\author[IPJ]{W.~Bardan}
\author[Edinb]{M.~Bashkanov}
\author[MS]{F.S.~Bergmann\corref{coau}}\ead{florianbergmann@uni-muenster.de}
\author[ASWarsH]{M.~Ber{\l}owski}
\author[Budker,Novosib]{A.~Bondar}
\author[PGI,DUS]{M.~B\"uscher}
\author[IKPUU]{H.~Cal\'{e}n}
\author[IFJ]{I.~Ciepa{\l}}
\author[PITue,Kepler]{H.~Clement}
\author[IPJ]{E.~Czerwi{\'n}ski}
\author[MS]{K.~Demmich}
\author[IKPJ]{R.~Engels}
\author[ZELJ]{A.~Erven}
\author[ZELJ]{W.~Erven}
\author[Erl]{W.~Eyrich}
\author[IKPJ,ITEP]{P.~Fedorets}
\author[Giess]{K.~F\"ohl}
\author[IKPUU]{K.~Fransson}
\author[IKPJ]{F.~Goldenbaum}
\author[IKPJ,IITI]{A.~Goswami}
\author[IKPJ,HepGat]{K.~Grigoryev}
\author[IKPUU]{C.--O.~Gullstr\"om}
\author[IKPJ,IASJ]{C.~Hanhart}
\author[IKPUU]{L.~Heijkenskj\"old\fnref{fnmz}}
\author[IKPJ]{V.~Hejny}
\author[MS]{N.~H\"usken}
\author[IPJ]{L.~Jarczyk}
\author[IKPUU]{T.~Johansson}
\author[IPJ]{B.~Kamys}
\author[ZELJ]{G.~Kemmerling\fnref{fnjcns}}
\author[IPJ]{G.~Khatri\fnref{fnharv}}
\author[MS]{A.~Khoukaz}
\author[IPJ]{A.~Khreptak}
\author[HeJINR]{D.A.~Kirillov}
\author[IPJ]{S.~Kistryn}
\author[ZELJ]{H.~Kleines\fnref{fnjcns}}
\author[Katow]{B.~K{\l}os}
\author[ASWarsH]{W.~Krzemie{\'n}}
\author[IFJ]{P.~Kulessa}
\author[IKPUU,ASWarsH]{A.~Kup{\'s}{\'c}}
\author[Budker,Novosib]{A.~Kuzmin}
\author[NITJ]{K.~Lalwani}
\author[IKPJ]{D.~Lersch}
\author[IKPJ]{B.~Lorentz}
\author[IPJ]{A.~Magiera}
\author[IKPJ,JARA]{R.~Maier}
\author[IKPUU]{P.~Marciniewski}
\author[ASWarsN]{B.~Maria{\'n}ski}
\author[ASWarsN]{H.--P.~Morsch}
\author[IPJ]{P.~Moskal}
\author[IKPJ]{H.~Ohm}
\author[IFJ]{W.~Parol}
\author[PITue,Kepler]{E.~Perez del Rio\fnref{fnlnf}}
\author[HeJINR]{N.M.~Piskunov}
\author[IKPJ]{D.~Prasuhn}
\author[IKPUU,ASWarsH]{D.~Pszczel}
\author[IFJ]{K.~Pysz}
\author[IKPUU,IPJ]{A.~Pyszniak}
\author[IKPJ,JARA,Bochum]{J.~Ritman}
\author[IITI]{A.~Roy}
\author[IPJ]{Z.~Rudy}
\author[IPJ]{O.~Rundel}
\author[IITB]{S.~Sawant}
\author[IKPJ]{S.~Schadmand}
\author[IPJ]{I.~Sch\"atti--Ozerianska}
\author[IKPJ]{T.~Sefzick}
\author[IKPJ]{V.~Serdyuk}
\author[Budker,Novosib]{B.~Shwartz}
\author[MS]{K.~Sitterberg}
\author[PITue,Kepler,Tomsk]{T.~Skorodko}
\author[IPJ]{M.~Skurzok}
\author[IPJ]{J.~Smyrski}
\author[ITEP]{V.~Sopov}
\author[IKPJ]{R.~Stassen}
\author[ASWarsH]{J.~Stepaniak}
\author[Katow]{E.~Stephan}
\author[IKPJ]{G.~Sterzenbach}
\author[IKPJ]{H.~Stockhorst}
\author[IKPJ,JARA]{H.~Str\"oher}
\author[IFJ]{A.~Szczurek}
\author[ASWarsN]{A.~Trzci{\'n}ski}
\author[IKPUU]{M.~Wolke}
\author[IPJ]{A.~Wro{\'n}ska}
\author[ZELJ]{P.~W\"ustner}
\author[KEK]{A.~Yamamoto}
\author[ASLodz]{J.~Zabierowski}
\author[IPJ]{M.J.~Zieli{\'n}ski}
\author[IKPUU]{J.~Z{\l}oma{\'n}czuk}
\author[ASWarsN]{P.~{\.Z}upra{\'n}ski}
\author[IKPJ]{M.~{\.Z}urek}
\author[IKPJ,IASJ]{\\[2ex] and\\[2ex] A.~Wirzba}

\address[IKPUU]{Division of Nuclear Physics, Department of Physics and
 Astronomy, Uppsala University, Box 516, 75120 Uppsala, Sweden}
\address[ASWarsN]{Department of Nuclear Physics, National Centre for Nuclear
 Research, ul.\ Hoza~69, 00-681, Warsaw, Poland}
\address[IPJ]{Institute of Physics, Jagiellonian University, prof.\
 Stanis{\l}awa {\L}ojasiewicza~11, 30-348 Krak\'{o}w, Poland}
\address[Edinb]{School of Physics and Astronomy, University of Edinburgh,
 James Clerk Maxwell Building, Peter Guthrie Tait Road, Edinburgh EH9 3FD,
 Great Britain}
\address[MS]{Institut f\"ur Kernphysik, Westf\"alische Wilhelms--Universit\"at
 M\"unster, Wilhelm--Klemm--Str.~9, 48149 M\"unster, Germany}
\address[ASWarsH]{High Energy Physics Department, National Centre for Nuclear
 Research, ul.\ Hoza~69, 00-681, Warsaw, Poland}
\address[Budker]{Budker Institute of Nuclear Physics of SB RAS, 11~akademika
 Lavrentieva prospect, Novosibirsk, 630090, Russia}
\address[Novosib]{Novosibirsk State University, 2~Pirogova Str., Novosibirsk,
 630090, Russia}
\address[PGI]{Peter Gr\"unberg Institut, PGI--6 Elektronische Eigenschaften,
 Forschungszentrum J\"ulich, 52425 J\"ulich, Germany}
\address[DUS]{Institut f\"ur Laser-- und Plasmaphysik, Heinrich--Heine
 Universit\"at D\"usseldorf, Universit\"atsstr.~1, 40225 D\"usseldorf, Germany}
\address[IFJ]{The Henryk Niewodnicza{\'n}ski Institute of Nuclear Physics,
 Polish Academy of Sciences, Radzikowskiego~152, 31--342 Krak\'{o}w, Poland}
\address[PITue]{Physikalisches Institut, Eberhard--Karls--Universit\"at
 T\"ubingen, Auf der Morgenstelle~14, 72076 T\"ubingen, Germany}
\address[Kepler]{Kepler Center f\"ur Astro-- und Teilchenphysik,
 Physikalisches Institut der Universit\"at T\"ubingen, Auf der
 Morgenstelle~14, 72076 T\"ubingen, Germany}
\address[IKPJ]{Institut f\"ur Kernphysik, Forschungszentrum J\"ulich, 52425
 J\"ulich, Germany}
\address[ZELJ]{Zentralinstitut f\"ur Engineering, Elektronik und Analytik,
 Forschungszentrum J\"ulich, 52425 J\"ulich, Germany}
\address[Erl]{Physikalisches Institut, Friedrich--Alexander--Universit\"at
 Erlangen--N\"urnberg, Erwin--Rommel-Str.~1, 91058 Erlangen, Germany}
\address[ITEP]{Institute for Theoretical and Experimental Physics named
 by A.I.\ Alikhanov of National Research Centre ``Kurchatov Institute'',
 25~Bolshaya Cheremushkinskaya, Moscow, 117218, Russia}
\address[Giess]{II.\ Physikalisches Institut, Justus--Liebig--Universit\"at
 Gie{\ss}en, Heinrich--Buff--Ring~16, 35392 Giessen, Germany}
\address[IITI]{Department of Physics, Indian Institute of Technology Indore,
 Khandwa Road, Simrol, Indore - 453552, Madhya Pradesh, India}
\address[HepGat]{High Energy Physics Division, Petersburg Nuclear Physics
 Institute named by B.P.\ Konstantinov of National Research Centre ``Kurchatov
 Institute'', 1~mkr.\ Orlova roshcha, Leningradskaya Oblast, Gatchina, 188300,
 Russia}
\address[IASJ]{Institute for Advanced Simulation, Forschungszentrum J\"ulich,
 52425 J\"ulich, Germany}
\address[HeJINR]{Veksler and Baldin Laboratory of High Energiy Physics,
 Joint Institute for Nuclear Physics, 6~Joliot--Curie, Dubna, 141980, Russia}
\address[Katow]{August Che{\l}kowski Institute of Physics, University of
 Silesia, Uniwersytecka~4, 40--007, Katowice, Poland}
\address[NITJ]{Department of Physics, Malaviya National Institute of
 Technology Jaipur, JLN Marg Jaipur - 302017, Rajasthan, India}
\address[JARA]{JARA--FAME, J\"ulich Aachen Research Alliance, Forschungszentrum
 J\"ulich, 52425 J\"ulich, and RWTH Aachen, 52056 Aachen, Germany}
\address[Bochum]{Institut f\"ur Experimentalphysik I, Ruhr--Universit\"at
 Bochum, Universit\"atsstr.~150, 44780 Bochum, Germany}
\address[IITB]{Department of Physics, Indian Institute of Technology Bombay,
 Powai, Mumbai - 400076, Maharashtra, India}
\address[Tomsk]{Department of Physics, Tomsk State University, 36~Lenina
 Avenue, Tomsk, 634050, Russia}
\address[KEK]{High Energy Accelerator Research Organisation KEK, Tsukuba,
 Ibaraki 305--0801, Japan}
\address[ASLodz]{Astrophysics Division, National Centre for Nuclear Research,
 Box~447, 90--950 {\L}\'{o}d\'{z}, Poland}
\address[IASJ]{Institute for Advanced Simulation and J\"ulich Center for Hadron Physics,
  Forschungszentrum J\"ulich, 52425, Germany}

\fntext[fnmz]{present address: Institut f\"ur Kernphysik, Johannes
 Gutenberg--Universit\"at Mainz, Johann--Joachim--Becher Weg~45, 55128 Mainz,
 Germany}
\fntext[fnjcns]{present address: J\"ulich Centre for Neutron Science JCNS,
 Forschungszentrum J\"ulich, 52425 J\"ulich, Germany}
\fntext[fnharv]{present address: Department of Physics, Harvard University,
 17~Oxford St., Cambridge, MA~02138, USA}
\fntext[fnlnf]{present address: INFN, Laboratori Nazionali di Frascati, Via
 E.~Fermi, 40, 00044 Frascati (Roma), Italy}

\cortext[coau]{Corresponding author }

\clearpage

\begin{abstract}
We report on the investigation of the rare decay $\eta\rightarrow\pi^0+\text{e}^++\text{e}^-$ 
which is of interest to study both $C$ violation in the electromagnetic interaction and to search for 
contributions from physics beyond the Standard Model, since the allowed 
decay via a two-photon intermediate state is strongly suppressed. 
The experiment has been performed using the WASA-at-COSY installation, located at the COSY accelerator 
of the Forschungszentrum J\"ulich, Germany. In total $3\times10^7$  events of the reaction
$\text{p}+\text{d}\rightarrow{^3\text{He}}+\eta$ have been recorded at an excess energy of 
$Q = 59.8\,\mathrm{MeV}$. Based on this data set the $C$ parity violating decay
$\eta\rightarrow\pi^0+\gamma^*\rightarrow\pi^0+\text{e}^++\text{e}^-$ via a
single-photon intermediate state has been searched for, resulting in new upper limits
of $\Gamma\left(\eta\rightarrow\pi^0+\text{e}^++\text{e}^-\right)/ 
\Gamma\left(\eta\rightarrow\pi^++\pi^-+\pi^0\right) < 3.28\times10^{-5}$
and 
$\Gamma\left(\eta\rightarrow\pi^0+\text{e}^++\text{e}^-\right)/ 
\Gamma\left(\eta\rightarrow\text{all}\right) < 7.5\times10^{-6}\ (\text{CL} = 90\,\%)$,
respectively.
\end{abstract}


\end{frontmatter}

\section{Introduction}
All strong and electromagnetic decays of the $\eta$ meson are either
suppressed or forbidden to first order. The $\eta$ meson is, in addition,
a $C$ and $P$ eigenstate of strong and electromagnetic interaction. This makes it
well suited for the study of rare processes and the search for forbidden ones.
The subject of this letter is the process
$\eta\rightarrow\pi^0+\text{e}^++\text{e}^-$ via the single-photon intermediate
state $\eta\rightarrow\pi^0+\gamma^*$ that would violate $C$ parity conservation.
The background for this process would be a two-photon process with an expected
branching ratio not larger than $10^{-8}$ according to theoretical calculations
\cite{Cheng1967,Smith1968,Ng1993}. The present experimental upper limit for the
branching ratio of the decay $\eta\rightarrow\pi^0+\text{e}^++\text{e}^-$ is
from the seventies of the last century and amounts only to $4.5\times10^{-5}$ (CL = $90\,\%$) \cite{Jane1975}.
A more stringent upper limit for
the decay channel $\eta\rightarrow\pi^0+\text{e}^++\text{e}^-$ has been determined
in the analysis presented in this paper.
The data have been collected using the WASA-at-COSY facility and also constituted the basis 
for studies of other $\eta$ meson decay channels already published in Ref.~\cite{Adlarson2016}.

\section{Experiment}
The WASA-at-COSY experiment was an internal experiment operated at the
accelerator COSY of the For\-schungs\-zentrum J\"ulich, Germany from 2006 to 2014 \cite{Hoistad2004}. 
For the measurements discussed here, a proton beam was accelerated to a
kinetic beam energy of $T_\text{p} = 1\,\mathrm{GeV}$ and collided with
deuterium pellets provided by the internal pellet target.
The $\eta$ mesons were produced in the reaction
$\text{p}+\text{d}\rightarrow{^3\text{He}}+\eta$.
\par
The WASA detector setup consists of two main parts: the central detector, which
was used for the reconstruction of the produced mesons and their decay
particles, and the forward detector used for the measurement of the
four momenta of the forward scattered $^3\text{He}$ nuclei.
A more detailed description of the WASA-at-COSY experimental setup can be found in
Ref.~\cite{Adlarson2016,Hoistad2004,Bargholtz2008}.
\par
The data for the studies presented here were obtained in two measurement periods, one
of four weeks in 2008 and one of eight weeks in 2009. For data acquisition
the trigger used required a large energy loss in subsequent scintillator
elements of the forward detector.
Since the $^3\text{He}$ nucleus stemming from the reaction
$\text{p}+\text{d}\rightarrow{^3\text{He}}+\eta$ is stopped in the first layer
of the WASA forward range hodoscope, a veto on the signals from the second layer
was used in addition. Due to the trigger relying on information from the forward detector
only, the utilized trigger was unbiased with respect to a decay
mode of the $\eta$ meson.
In total about $3\times10^7$ events containing an $\eta$ meson were recorded
with $1\times10^7$ events originating from the 2008 period and $2\times10^7$ events from the
2009 period \cite{Adlarson2016}.

\section{Data analysis}
The analysis of the decay $\eta\rightarrow\pi^0+\text{e}^++\text{e}^-$
was based on a common analysis chain for $\eta$ decay studies described in
Ref.~\cite{Adlarson2016}. Since only very few events were expected to remain in
the analysis after the event selection, an optimal choice of the selection
conditions is important for the best possible result. These conditions
were determined with the aid of an optimization algorithm based on Monte
Carlo simulations.

\paragraph{Preselection}
Before the selection conditions for the decay
$\eta\rightarrow\pi^0+\text{e}^++\text{e}^-$ were determined, the data collected
in 2008 and 2009 were preselected with conditions common to all recorded
reactions. For instance, conditions on time correlations of the
measured particles were used, as presented in Ref.~\cite{Adlarson2016}.
Furthermore, to reject hits from particles that were wrongly identified as
secondary particles (so-called split-offs) and
electron-positron pairs from conversion of photons at the COSY beam pipe,
two-dimensional cuts were utilized. More details of these
conditions were published in Ref.~\cite{Adlarson2016}.
\par
Besides these general preselection conditions, a cut on the signature of the
decay $\eta\rightarrow\pi^0+\text{e}^++\text{e}^-$ was applied requesting at
least one positively and one negatively charged particle detected in the central
detector, as well as at least two neutral particles originating from the
$\pi^0$ meson decay $\pi^0\rightarrow\gamma+\gamma$. The last condition
applied for data preselection requires the maximum considered momenta of the 
charged decay particles to be below
$p = 250\,\mathrm{MeV}/c$, since the momenta of the leptons
of the decay $\eta\rightarrow\pi^0+\text{e}^++\text{e}^-$ are expected to be
below this value.

\paragraph{Monte Carlo simulations}
In order to determine optimal selection conditions for the search for the decay
channel $\eta\rightarrow\pi^0+\text{e}^++\text{e}^-$, $1.8\times10^8$ 
Monte-Carlo events of all non-signal $\eta$ decays observed yet were created
with respect to their relative branching ratio \cite{Patrignani2016}, as well as
two million events for the signal decay. These simulations were generated with the
\textsc{pluto++} software package \cite{Frohlich2007} considering the angular 
distribution of $\text{p}+\text{d}\rightarrow{^3\text{He}}+\eta$ at
$T_\text{p} = 1\,\mathrm{GeV}$ according to Ref.~\cite{Adlarson2014}. For the
various $\eta$ decay channels physics models as included in \textsc{pluto++}
were used. The reader is refered to Ref.~\cite{Adlarson2016} for further details.
\par
In addition to the simulations of $\eta$ decays, about $4.3\times10^9$ events
for the direct pion production were created, with most events for the
production reactions $\text{p}+\text{d}\rightarrow{^3\text{He}}+\pi^0+\pi^0$
and $\text{p}+\text{d}\rightarrow{^3\text{He}}+\pi^++\pi^-$,
as these contribute most to the non-$\eta$ background at the given kinetic
beam energy. For these two-pion productions the ABC effect was incorporated
into the simulations according to the model discussed in Ref.~\cite{Adlarson2015}.
\par
\begin{figure}
	\centering
	\resizebox{0.45\textwidth}{!}{
  \includegraphics{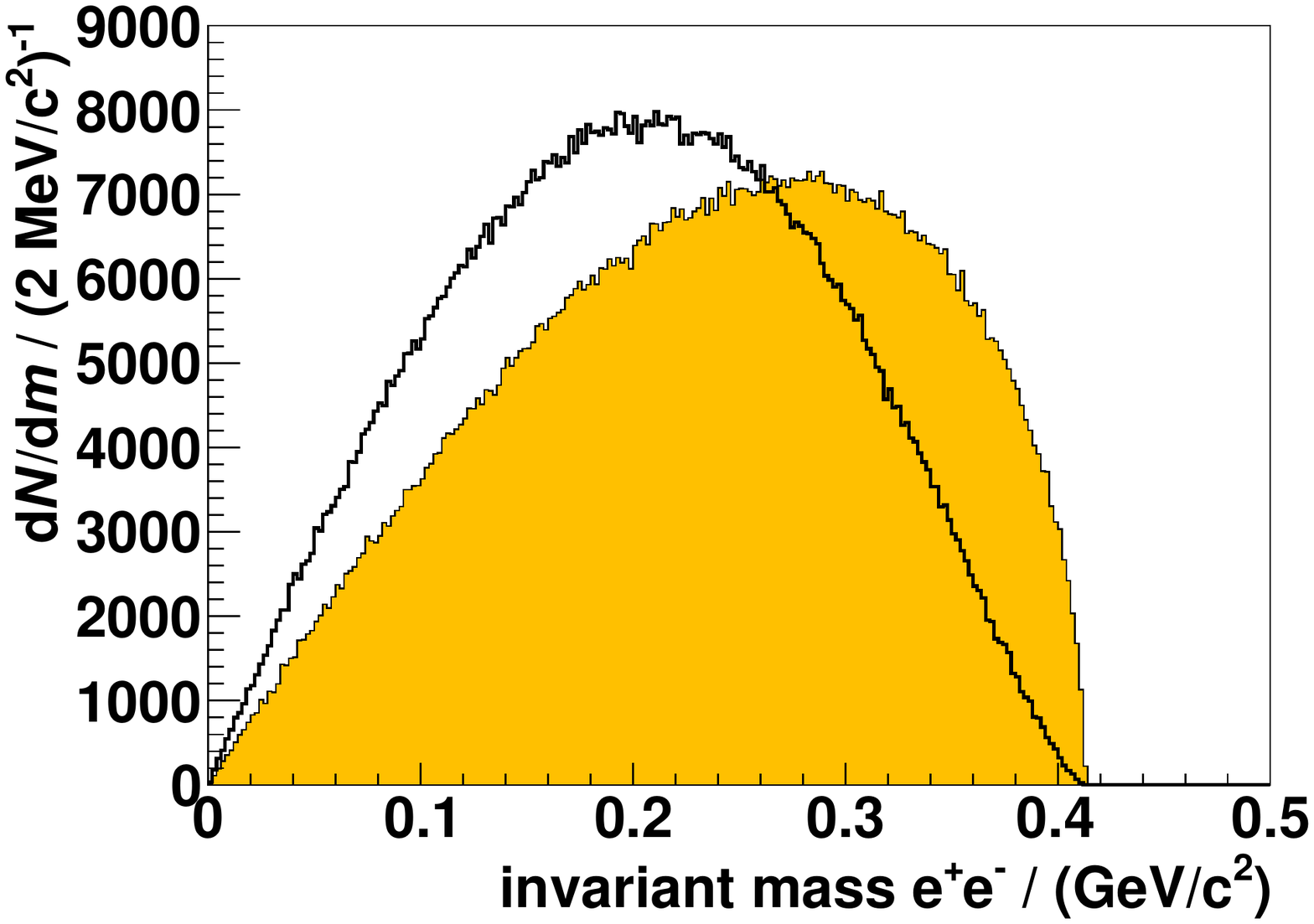}
	}
	\caption{Invariant mass of $\text{e}^+\text{e}^-$ pairs for the simulated
	decay $\eta\rightarrow\pi^0+\text{e}^++\text{e}^-$.
	Black lined: decay via $\eta\rightarrow\pi^0+\gamma^*$ considering VMD.
	Shadowed in orange: decay according to three-particle phase space.}
	\label{fig:etapi0eeIMee}
\end{figure}

The simulations for the signal decay $\eta\rightarrow\pi^0+\text{e}^++\text{e}^-$
were generated with two different model assumptions. The first one is a
decay according to pure three-particle phase space. The second is based on the
vector meson dominance (VMD) model for the intermediate virtual photon. The
direct decay $\eta\rightarrow\pi^0+\gamma$ violates both $C$ parity and angular
momentum conservation plus global gauge invariance.
Thus, there is no $\eta\rightarrow\pi^0+\gamma$ on-shell
contribution for the decay $\eta\rightarrow\pi^0+\text{e}^++\text{e}^-$
and the transition form factor for the off-shell contribution
vanishes at zero virtuality, such that the single-photon pole is
completely removed \cite{Bernstein1965,Barrett1966,Bazin1968}.
In Fig.~\ref{fig:etapi0eeIMee} the invariant mass of the $\text{e}^+\text{e}^-$
pair produced in the decay is plotted according to three-particle
phase space (shadowed in orange) and the decay via $\eta\rightarrow\pi^0+\gamma^*$
according to the discussed model. A more detailed
calculation of the model can be found in Ref.~\cite{Bergmann2017}.
\par
\par
To simulate the WASA detector responses, the WASA Monte Carlo package \textsc{wmc}
was used, which is based on \textsc{geant3} \cite{Geant1994}. The settings for
the spatial, timing and energy resolution in \textsc{wmc} were set to agree
with the resolution observed in data.
\par
Due to the high luminosities of the WASA-at-COSY experiment, it is possible that
detector responses from one event can overlap with another event. This effect
was considered in the simulations and the amount of event overlap was left as a
free parameter for the fit of the simulations to data (see next paragraph).
\par
All Monte Carlo simulations were preselected with conditions identical to those
for data preselection.

\paragraph{Data description}
The choice of the selection conditions with regard to the decay channel
$\eta\rightarrow\pi^0+\text{e}^++\text{e}^-$ is based on Monte Carlo simulations.
It is necessary to know the contributions of the various reactions to the
collected data for an optimal choice. Therefore, the
2008 and 2009 data sets were fitted separately in distributions of selected
quantities by template distributions of the aforementioned Monte Carlo simulations
to determine the contributions of the individual reactions to the data.
In detail, these distributions are:
\begin{itemize}
	\item the missing mass $m_\text{X}$, corresponding to the invariant mass
	of the proton beam and the deuteron target remaining after the $^3\text{He}$
	four momentum has been subtracted and peaks at the $\eta$ mass for the
	reaction $\text{p}+\text{d}\rightarrow{^3\text{He}}+\eta$,
	\item the invariant mass $m_{\text{e}\text{e}\gamma\gamma}$ of an
	electron-positron pair candidate and two photons, which peaks at the $\eta$
	mass for the decay $\eta\rightarrow\pi^0+\text{e}^++\text{e}^-$ with
	$\pi^0\rightarrow\gamma+\gamma$,
	\item the invariant mass $m_{\gamma\gamma}$ of two photons, which peaks at
	the $\pi^0$ mass for reactions with $\pi^0$ mesons produced,
	\item the invariant mass $m_{\text{e}\text{e}}$ of an electron-positron pair
	candidate,
	\item the smallest invariant mass $m_{\text{e}\gamma}$ of all four possible
	combinations of an electron or positron candidate and a photon and
	\item the missing mass squared $m_\text{Xee}^2$, which is the invariant mass
	squared of the proton beam and the deuteron target remaining after the
	$^3\text{He}$ four momentum and the electron-positron pair candidate momentum
	have been subtracted and peaks at the $\pi^0$ mass squared for the reaction of
	interest.	
\end{itemize}
\par
Under the assumption of a branching ratio of the decay below
the current upper limit of $4.5\times10^{-5}$ (CL = $90\,\%$) \cite{Jane1975},
there are less than 150 events expected from the decay
$\eta\rightarrow\pi^0+\text{e}^++\text{e}^-$ in the combined data sets after
preselection, considering the preselection efficiency for the signal decay.
A fit by Monte Carlo simulations including the simulated decay
$\eta\rightarrow\pi^0+\text{e}^++\text{e}^-$ is consistent with zero events from
this signal decay channel. Therefore, the decay
$\eta\rightarrow\pi^0+\text{e}^++\text{e}^-$ was excluded from the fit.
While the differential
distribution for the reaction $\text{p}+\text{d}\rightarrow{^3\text{He}}+\eta$
is well known \cite{Adlarson2014}, the differential distributions are known only with high
uncertainties or not at all for direct multi-pion productions. Hence, the data
were divided into ten bins in angular ranges of $\cos{\vartheta_{^3\text{He}}^\text{cms}}$\footnote{
$\vartheta_{^3\text{He}}^\text{cms}$ is the polar scattering angle of the
$^3\text{He}$ nucleus relative to the beam axis in the center of mass system.}.
Monte Carlo simulations were fitted to data in the eight angular bins ranging
from $-1$ to $0.6$. The angular range
$0.6 < \cos{\vartheta_{^3\text{He}}^\text{cms}} \leq 1$ was excluded because of
the lower energy resolution of the forward detector for these forward scattered
$^3\text{He}$ nuclei. Moreover, the relative amount of background from the
direct pion production is larger in this angular range, whereas less than
$3\,\%$ of all $\text{p}+\text{d}\rightarrow{^3\text{He}}+\eta$ events have a
$\cos{\vartheta_{^3\text{He}}^\text{cms}} > 0.6$.

\par
\begin{figure}
	\centering
	\resizebox{0.45\textwidth}{!}{
  \includegraphics{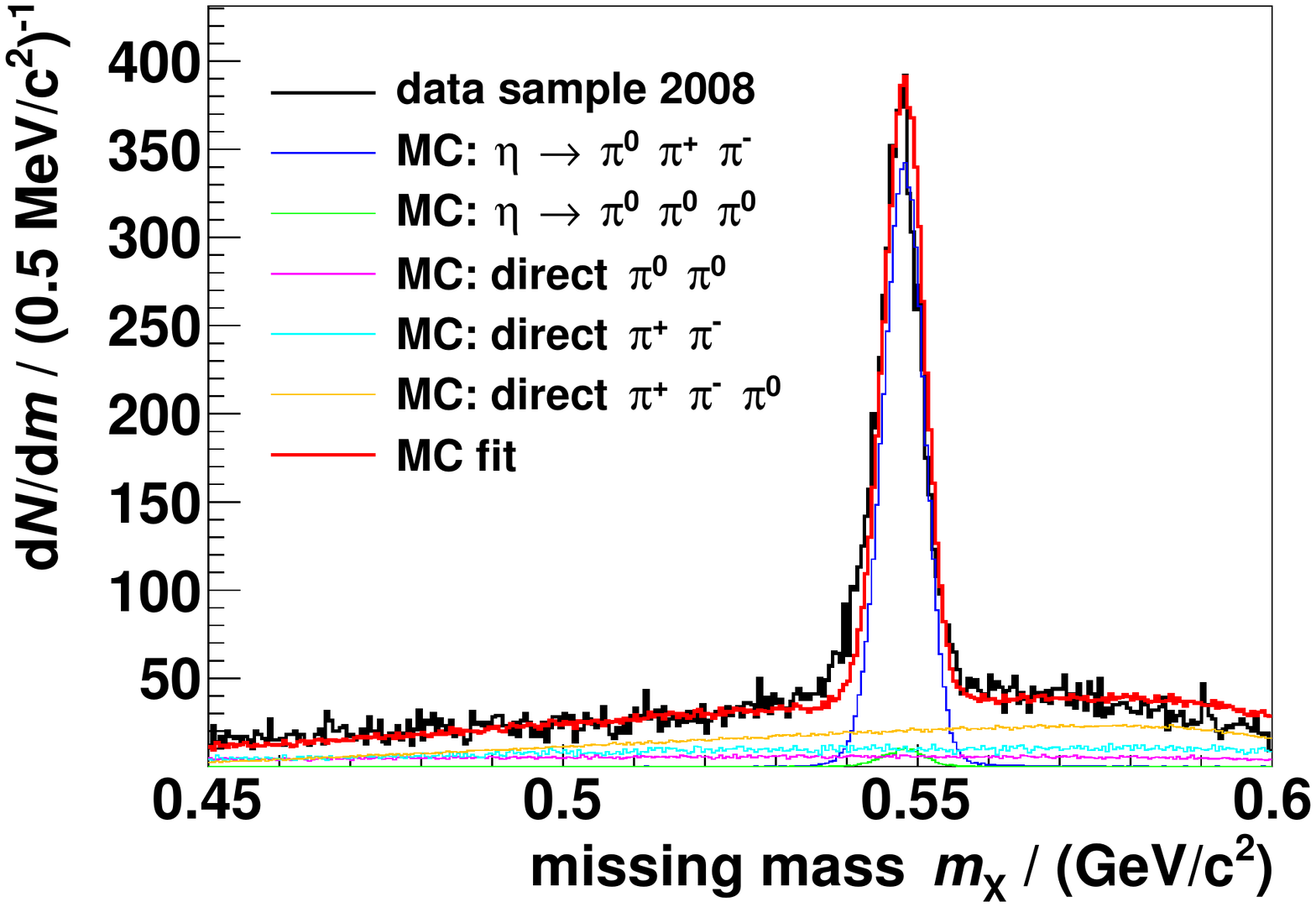}
	}
	\caption{Missing mass
	$m_\text{X} = \left|\mathbb{P}_\text{p}+\mathbb{P}_\text{d}-\mathbb{P}_{^3\text{He}}\right|$
	after preselection for a data sample of the 2008 period fitted by Monte Carlo
	simulations. Only the most common contributions of the various reactions to
	the fit are plotted separately.}
	\label{fig:DataFit_mX}
\end{figure}
\par
\begin{figure}
	\centering
	\resizebox{0.45\textwidth}{!}{
  \includegraphics{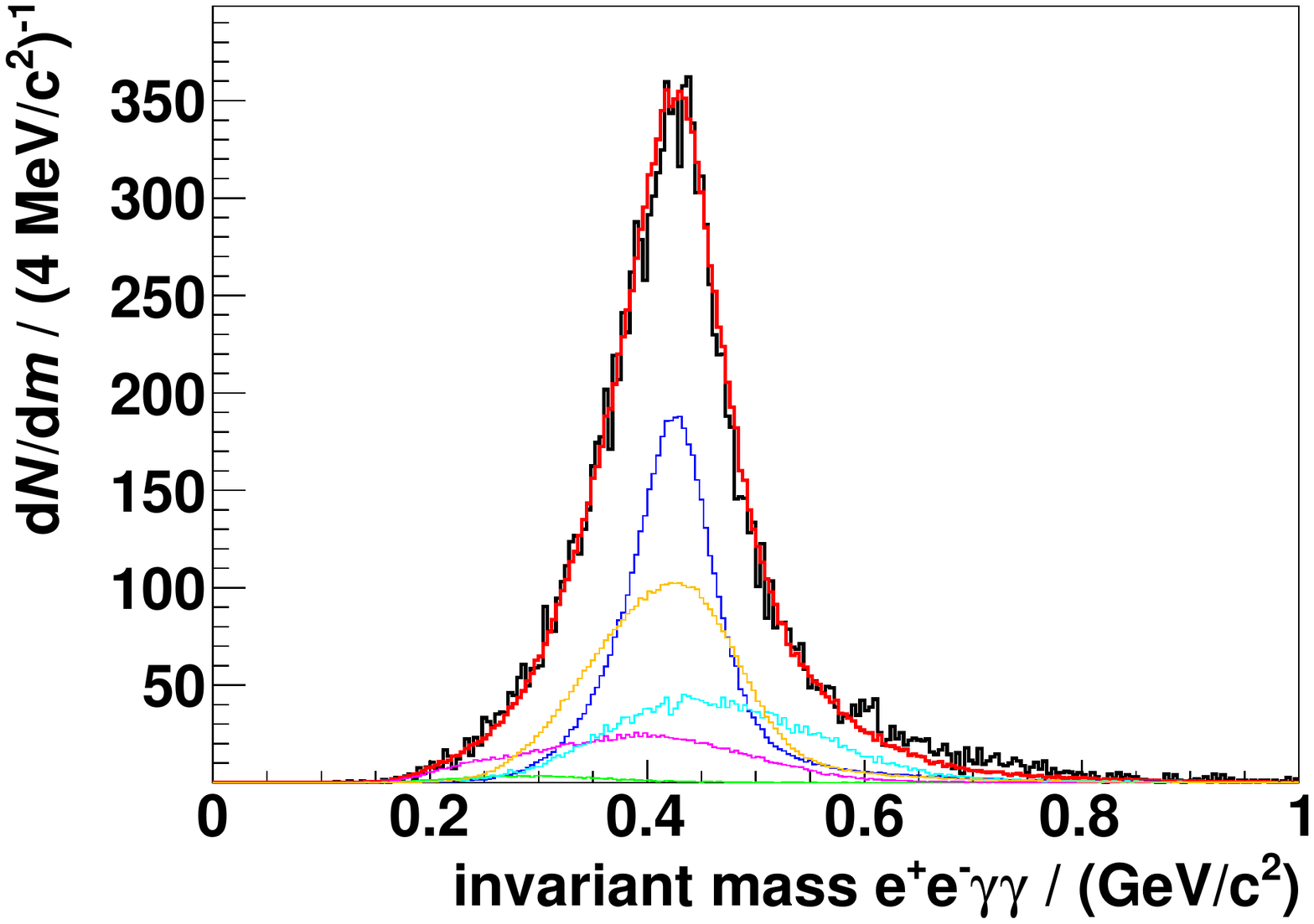}
	}
	\caption{Invariant mass of $\text{e}^+\text{e}^-\gamma\gamma$
	after preselection for a data sample of the 2008 period fitted by
	Monte Carlo simulations. For the legend see Fig.~\ref{fig:DataFit_mX}.}
	\label{fig:DataFit_meegg}
\end{figure}
\par
\begin{figure}
	\centering
	\resizebox{0.45\textwidth}{!}{
  \includegraphics{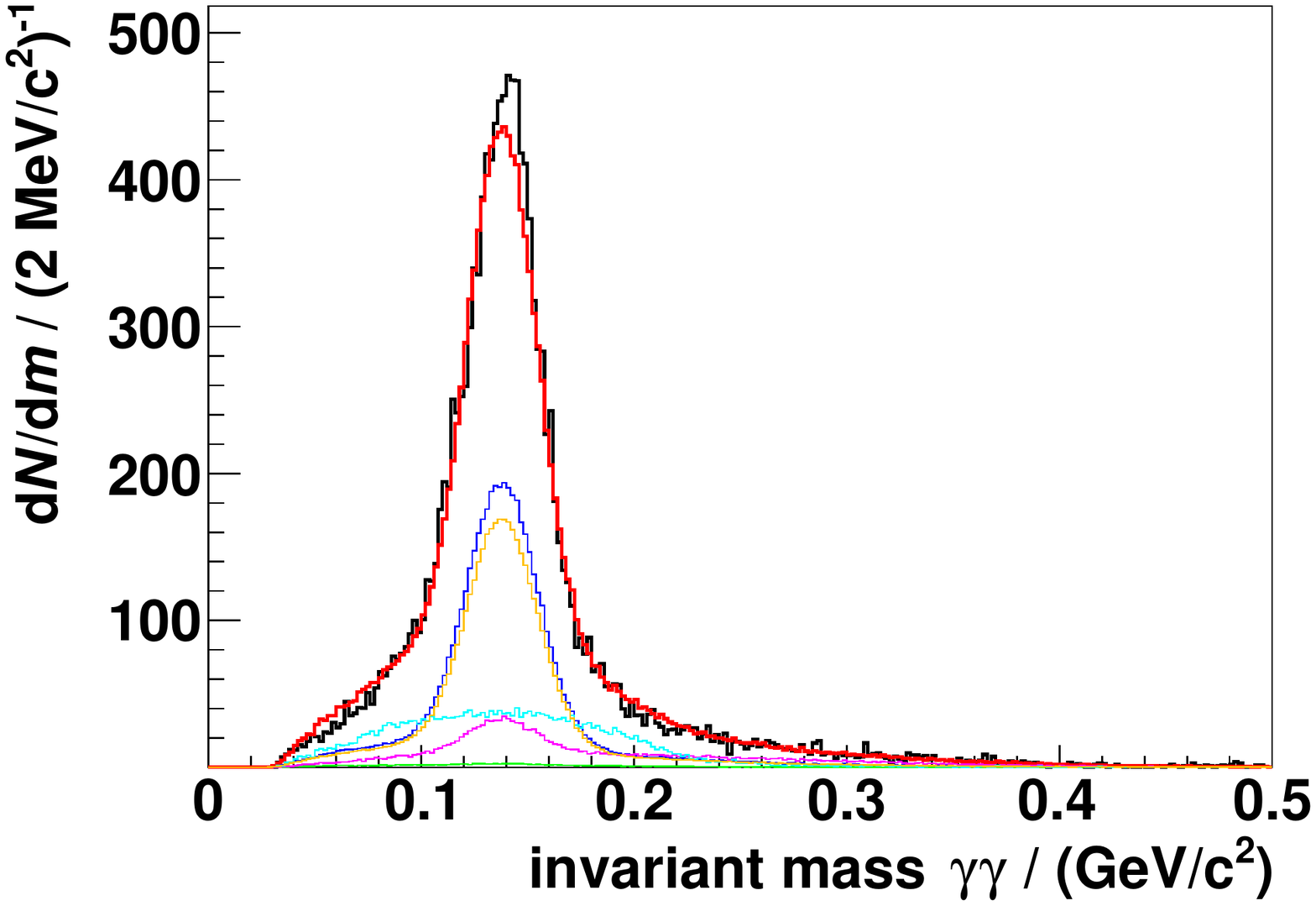}
	}
	\caption{Invariant mass of $\gamma\gamma$
	after preselection for a data sample of the 2008 period fitted by
	Monte Carlo simulations. For the legend see Fig.~\ref{fig:DataFit_mX}.}
	\label{fig:DataFit_mgg}
\end{figure}
\par
\begin{figure}
	\centering
	\resizebox{0.45\textwidth}{!}{
  \includegraphics{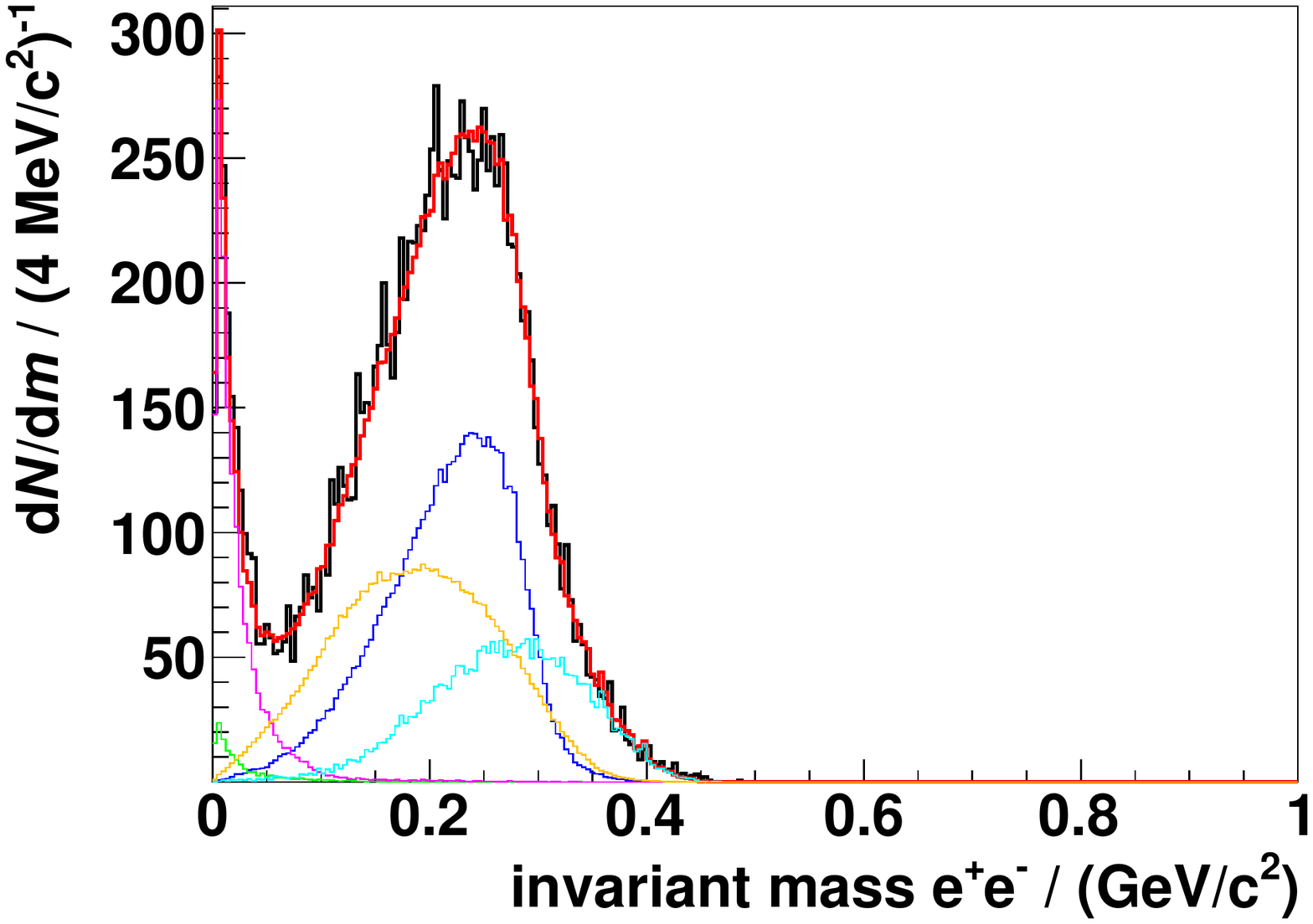}
	}
	\caption{Invariant mass of $\text{e}^+\text{e}^-$
	after preselection for a data sample of the 2008 period fitted by
	Monte Carlo simulations. For the legend see Fig.~\ref{fig:DataFit_mX}.}
	\label{fig:DataFit_mee}
\end{figure}
\par
The fit of the Monte Carlo simulations to the data was performed simultaneously
for all angular ranges and distributions with identical scaling parameters for the
simulations for all distributions within one angular range. Furthermore, the ratios for the various $\eta$
decays were constrained to the branching ratios according to
Ref.~\cite{Patrignani2016} within the given uncertainties. These were set to be
identical for all angular ranges. Similarly, the amount of event overlap was
included as one global fit parameter. In Fig.~\ref{fig:DataFit_mX},
Fig.~\ref{fig:DataFit_meegg}, Fig.~\ref{fig:DataFit_mgg} and
Fig.~\ref{fig:DataFit_mee} the resulting Monte Carlo fits to the 2008 data
are plotted for $m_\text{X}$, $m_{\text{e}\text{e}\gamma\gamma}$,
$m_{\gamma\gamma}$ and $m_{\text{e}\text{e}}$ for the angular range
$0.2 < \cos{\vartheta_{^3\text{He}}^\text{cms}} \leq 0.4$. According to this fit
most events remaining after preselection originate from the $\eta$ decay
$\eta\rightarrow\pi^++\pi^-+\pi^0$, the direct
$\text{p}+\text{d}\rightarrow{^3\text{He}}+\pi^++\pi^-+\pi^0$ production
and the direct two-pion production reactions. A collection of all fits is
available in Ref.~\cite{Bergmann2017}.

\paragraph{Selection conditions}
The selection conditions for the search for the decay
$\eta\rightarrow\pi^0+\text{e}^++\text{e}^-$ were based on the following quantities:
\begin{itemize}
	\item the missing mass $m_\text{X}$ to identify the production
	reaction $\text{p}+\text{d}\rightarrow{^3\text{He}}+\eta$,
	\item the invariant mass $m_{\text{e}\text{e}\gamma\gamma}$ of an
	electron-positron pair candidate and two photons to select the decay
	$\eta\rightarrow\pi^0+\text{e}^++\text{e}^-\rightarrow\gamma+\gamma+\text{e}^++\text{e}^-$,
	\item the invariant mass $m_{\gamma\gamma}$ of two photons to ascertain the
	decay $\pi^0\rightarrow\gamma+\gamma$,
	\item the invariant mass $m_{\text{e}\text{e}}$ of an electron-positron pair
	candidate,
	\item the $\chi^2$ probability of a kinematic fit with the hypothesis
	$\text{p}+\text{d}\rightarrow{^3\text{He}}+\gamma+\gamma+\text{e}^++\text{e}^-$
	and
	\item the energy loss $E_\text{dep}^\text{SEC}$ of the charged particles in
	the central detector scintillator electromagnetic calorimeter (SEC) and their momentum $p$ to discriminate
	$\text{e}^\pm$ and $\pi^\pm$ (particle identification, PID).
\end{itemize}
\par
\begin{figure}
	\centering
	\resizebox{0.45\textwidth}{!}{
  \includegraphics{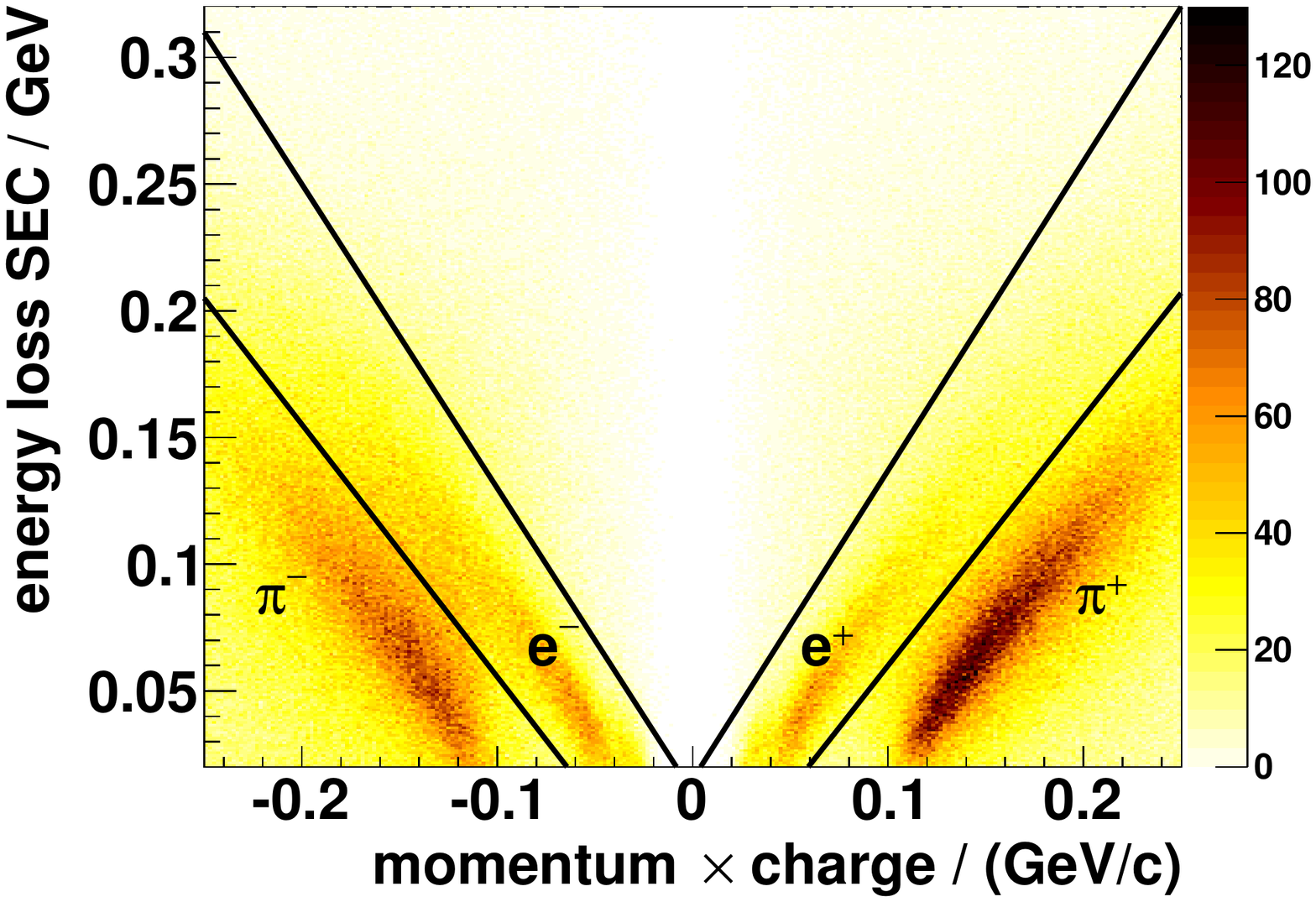}
	}
	\caption{Energy loss of charged particles in the SEC plotted against their
	momentum times charge for the preselected data sets. A graphical cut
	around the electron and positron band is indicated by black lines.}
	\label{fig:PID2009}
\end{figure}
\par
The choice of the cut conditions was performed with $40\,\%$ of the generated
Monte Carlo simulations, where\-as the remaining Monte Carlo data sample was used
later for the selection efficiency determination. While the graphical cut for the particle identification
was chosen beforehand (see Fig.~\ref{fig:PID2009}), as it is a common cut utilized
for PID independent from the analyzed reaction, the selection conditions for
the other five quantities were determined by an optimization algorithm.
This algorithm is based on the relative amount of simulated signal events
$S_\text{R} = N_\text{S}^\text{cut} / N_\text{S}^\text{pres}$ remaining after
all cuts ($N_\text{S}^\text{cut}$) compared to the number after preselection
($N_\text{S}^\text{pres}$) and the relative amount of all simulated background
events $B_\text{R} = N_\text{B}^\text{cut} / N_\text{B}^\text{pres}$ remaining
after all cuts ($N_\text{B}^\text{cut}$) in relation to the number after
preselection ($N_\text{B}^\text{pres}$). In case of the background reactions the
contributions as obtained in the data description were used to downscale the Monte
Carlo simulations and to extract the numbers.
\par
The cut optimization algorithm maximizes the evaluation function
\begin{equation}
\label{eq:evaluationfunction}
	G = S_\text{R} \cdot \frac{S_\text{R}}{B_\text{R}}
\end{equation}
by varying the selection conditions for all chosen quantities.
\par
With the aid of the cut optimization algorithm the following selection
conditions were determined:
\begin{align}
\label{eq:selection}
	0.5414\,\text{GeV}/c^2 &\leq &&m_\text{X} &&\leq 0.5561\,\text{GeV}/c^2\,, \\
	0.507\,\text{GeV}/c^2 &\leq &&m_{\text{ee}\gamma\gamma} &&\leq 0.646\,\text{GeV}/c^2\,, \\
	0.0923\,\text{GeV}/c^2 &\leq &&m_{\gamma\gamma} &&\leq 0.1574\,\text{GeV}/c^2\,, \\
	& &&m_\text{ee} &&\geq 0.096\,\text{GeV}/c^2\,\text{ and} \\
	& &&\chi^2 \text{prob.} &&\geq 0.05\,.
\end{align}

\section{Results}
\begin{figure}
	\centering
	\resizebox{0.45\textwidth}{!}{
  \includegraphics{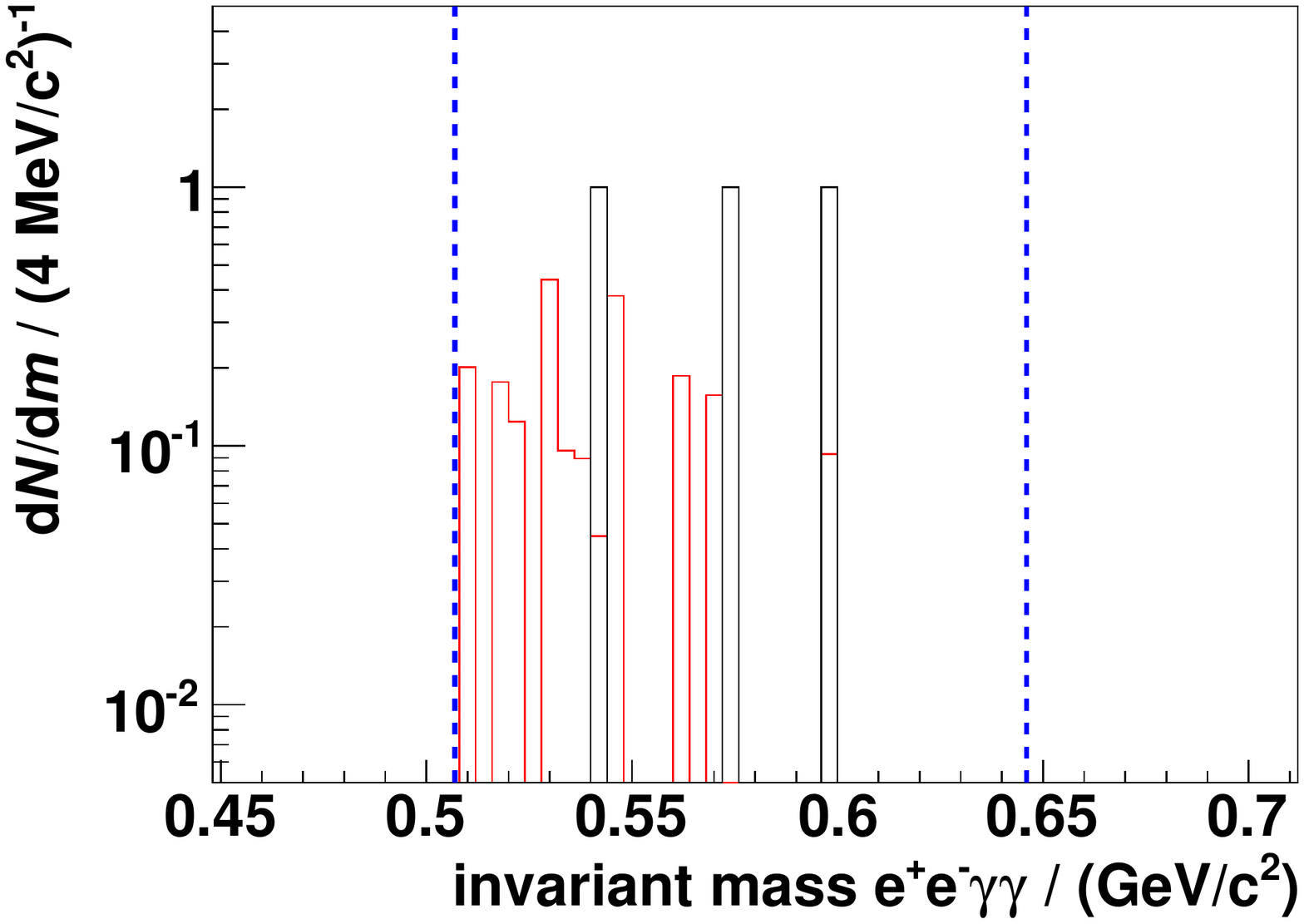}
	}
	\caption{Invariant mass of $\text{e}^+\text{e}^-\gamma\gamma$ after all
	cuts for the 2008 and 2009 data sets (black) and for the simulations scaled to
	data according to the fit to data after preselection (red). The blue dashed
	lines indicate the chosen selection conditions.}
	\label{fig:Resultmeegg}
\end{figure}

After applying the selection conditions to the data, three
events were left, whereas two events were expected to remain from the direct two-pion
production $\text{p}+\text{d}\rightarrow{^3\text{He}}+\pi^0+\pi^0$ according to
Monte Carlo simulations. All other background reaction channels were found to give 
no sizeable contribution after applying the cuts.
The invariant mass, $m_{\text{ee}\gamma\gamma}$, for these events are plotted in
Fig.~\ref{fig:Resultmeegg} together with simulated data.
Note that the generated
Monte Carlo events were scaled according to the fit to data after
preselection and that the sum of all Monte Carlo events remaining after all cuts
is equal to two events.
\par
\par
The overall reconstruction efficiency for the signal decay
$\eta\rightarrow\pi^0+\text{e}^++\text{e}^-$ was determined to be
\begin{equation}
\label{eq:effvitual}
	\varepsilon_\text{S}^\text{virtual} = 0.02331(7)
\end{equation}
for a decay via $\eta\rightarrow\pi^0+\gamma^*$ assuming VMD,
whereas the assumption of a decay according to pure three-particle phase space
results in
\begin{equation}
\label{eq:effphase}
	\varepsilon_\text{S}^\text{phase} = 0.01844(7).
\end{equation}
The given uncertainties are purely statistical ones.
\par
In order to calculate the upper limit for the branching ratio
$\Gamma(\eta\rightarrow\pi^0+\text{e}^++\text{e}^-) / \Gamma(\eta\rightarrow\text{all})$,
the decay channel $\eta\rightarrow\pi^++\pi^-+\pi^0$ with $\pi^0\rightarrow\gamma+\gamma$
was utilized for normalization. This is a reasonable choice as this decay
channel has the same signature as the signal decay and, thus, possible systematic
effects introduced by differences of the signature are avoided.
According to the data description by Monte Carlo
simulations there were
\begin{equation}
\label{eq:netaprodced}
	N_{\eta\rightarrow\pi^+\pi^-\pi^0_{\gamma\gamma}}^\text{produced} = (6.509 \pm 0.018) \times 10^6
\end{equation}
events in data, considering already the efficiency correction
determined by Monte Carlo studies. In order to determine a final upper limit for
the branching ratio of $\eta\rightarrow\pi^0+\text{e}^++\text{e}^-$,
all uncertainties have to be considered and incorporated into the calculations.

\paragraph{Systematics}
The systematic and statistical uncertainties, which need to be considered for
the upper limit determination, can be separated into uncertainties by
multiplicative effects and uncertainties by offset effects. The former include
an uncertainty of the reconstruction efficiency of the decay
$\eta\rightarrow\pi^0+\text{e}^++\text{e}^-$ and an uncertainty in the number of
$\eta\rightarrow\pi^++\pi^-+\left(\pi^0\rightarrow\gamma+\gamma\right)$ events
in data. The latter ones are uncertainties of the number of background events
remaining after all cuts.
\par
To determine the systematic uncertainty for the signal reconstruction efficiency,
the resolution settings for the Monte Carlo simulations were varied within the
uncertainties of the individual detector resolutions observed in data.
The extracted square root of the relative variance of the reconstruction
efficiency was found to be
\begin{equation}
\label{eq:varvirtual}
	\sqrt{\text{Var}_\text{rel}^\text{virtual}} = 0.059
\end{equation}
for a decay via $\eta\rightarrow\pi^0+\gamma^*$ assuming VMD whereas
for a decay according to pure three-particle phase space one finds 
\begin{equation}
\label{eq:varphase}
	\sqrt{\text{Var}_\text{rel}^\text{phase}} = 0.057.
\end{equation}
In the following analysis the square root of the variance was considered as
the systematic uncertainty.
\par
The uncertainty for the efficiency corrected number of
$\eta\rightarrow\pi^++\pi^-+\left(\pi^0\rightarrow\gamma+\gamma\right)$ events
in data was obtained by a comparison to the efficiency corrected number
determined utilizing less strict preselection conditions, namely no cuts to
reject conversion or split-off events, no cut on the momentum of charged decay
particles and less strict cuts on the particles' energies. Hereby a systematic
uncertainty of $2.3\,\%$ was determined.
\par
\begin{figure}
	\centering
	\resizebox{0.45\textwidth}{!}{
  \includegraphics{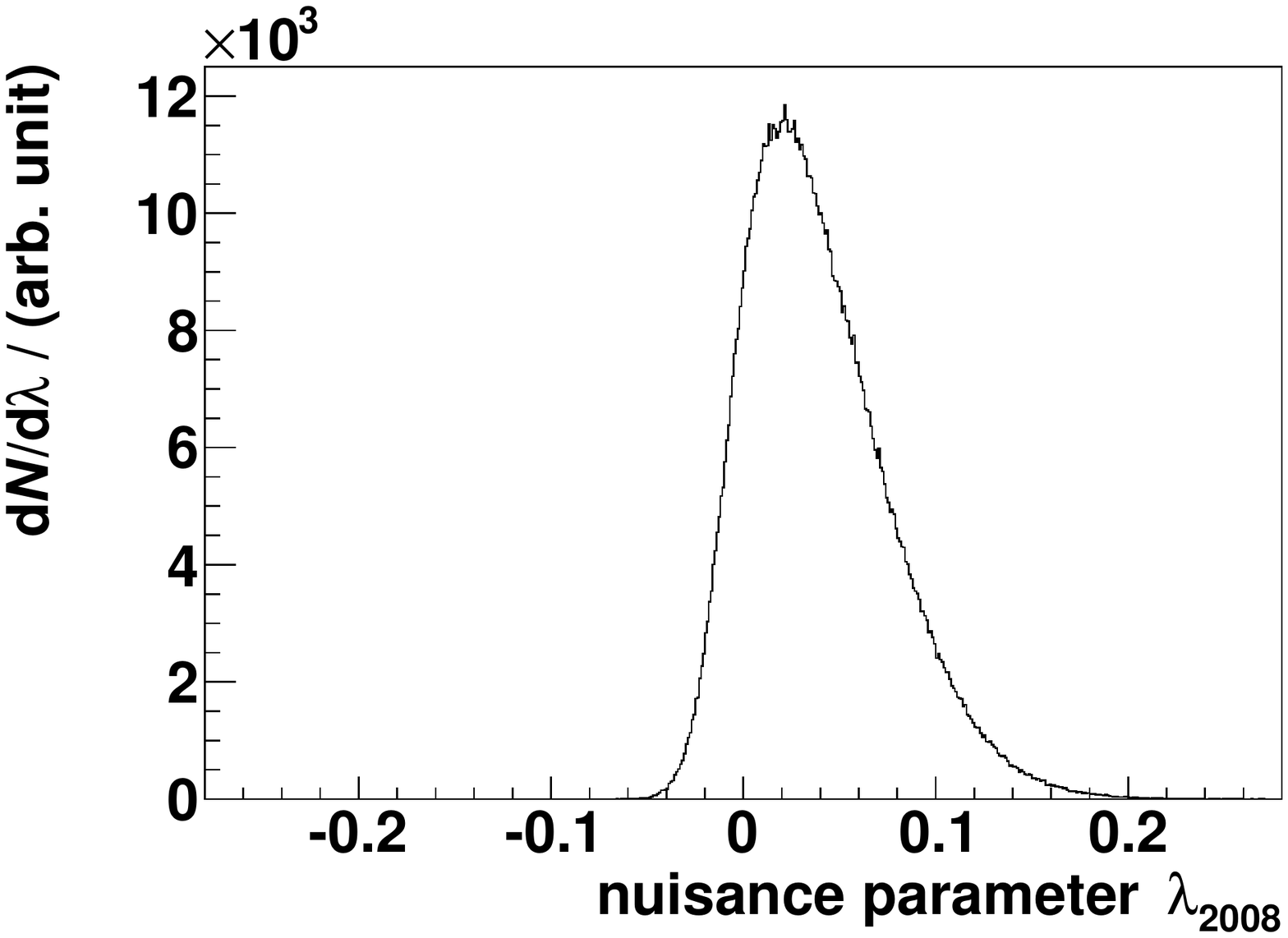}
	}
	\caption{Nuisance parameter $\lambda_{2008}$ for the systematic uncertainty
	of the number of background events remaining after all cuts in the 2008 data
	set.}
	\label{fig:Nuisance}
\end{figure}

The uncertainties for the number of background events remaining after all cuts
can be separated into a statistical uncertainty due to the finite number of
Monte Carlo simulations and systematic uncertainties introduced by uncertainties
of the fit of Monte Carlo simulations to data. The latter are dominated by
differences between the Monte Carlo fit parameters for the 2008 and 2009 data
sets, leading to asymmetric uncertainties. Such different fit parameters 
for both data sets originated mainly from different experimental settings, which
affected, e.g., the event overlap due to different luminosities. To determine
the overall systematic uncertainty for the number of remaining background
events, the probability density functions (pdf) of the individual uncertainties
were folded. The resulting pdf for the nuisance parameters $\lambda_{2008}$ and
$\lambda_{2009}$ corresponds to the overall relative systematic uncertainty for
the 2008 and 2009 data sets and was incorporated into the upper limit
calculations. In Fig.~\ref{fig:Nuisance} the distribution of the nuisance
parameter is illustrated for the 2008 data set.
\par
\par
In order to investigate further possible systematic effects, the selection
conditions used for the analysis were varied and the expectations according to
simulations were compared to the number of events seen in data. Since the
expected number of events agreed with the number of events seen in data within
the statistical uncertainties, no additional systematic effect needs to be
considered.
\par
A detailed description of the uncertainty investigations is available in
Ref.~\cite{Bergmann2017}.

\paragraph{Upper limit}
The upper limit for the relative branching ratio of the decay
$\eta\rightarrow\pi^0+\text{e}^++\text{e}^-$ was calculated with the formula:
\begin{equation}
\label{eq:brformula}
	\frac{\Gamma\left(\eta\rightarrow\pi^0+\text{e}^++\text{e}^-\right)}
	{\Gamma\left(\eta\rightarrow\pi^++\pi^-+\pi^0\right)}
	< \frac{N_\text{S,up}}{N_{\eta\rightarrow\pi^+\pi^-\pi^0}^\text{produced} \cdot \varepsilon_\text{S}}
\end{equation}
with the upper limit $N_\text{S,up}$ for the number of signal events, which
depends on the number of observed events and the number of expected
background events. For the calculation of $N_\text{S,up}$ a Bayesian approach
was chosen as given in Ref.~\cite{Zhu2007} with a flat prior pdf and incorporating
the determined uncertainties and the pdfs for the nuisance parameters.
\par
As a result the relative branching ratio of the decay $\eta\rightarrow\pi^0+\text{e}^++\text{e}^-$ 
via $\eta\rightarrow\pi^0+\gamma^*$ and assuming VMD was found to be 
\begin{align}
\label{eq:brrelvirtual}
	\frac{\Gamma\left(\eta\rightarrow\pi^0+\text{e}^++\text{e}^-\right)_\text{virtual}}
	{\Gamma\left(\eta\rightarrow\pi^++\pi^-+\pi^0\right)}
	&< 3.28\times10^{-5} \nonumber \\
	&\qquad (\text{CL} = 90\,\%)
\end{align}
whereas the assumption of a pure three-particle phase space distribution of the 
ejectiles results in 
\begin{align}
\label{eq:brrelphase}
	\frac{\Gamma\left(\eta\rightarrow\pi^0+\text{e}^++\text{e}^-\right)_\text{phase}}
	{\Gamma\left(\eta\rightarrow\pi^++\pi^-+\pi^0\right)}
	&< 4.14\times10^{-5} \nonumber \\
	&\qquad (\text{CL} = 90\,\%).
\end{align}
Considering the branching ratio of the decay
$\eta\rightarrow\pi^++\pi^-+\pi^0$ of
$\Gamma\left(\eta\rightarrow\pi^++\pi^-+\pi^0\right) /
\Gamma\left(\eta\rightarrow\text{all}\right) = 0.2292(28)$ \cite{Patrignani2016},
the new upper limit for the branching ratio of the decay
$\eta\rightarrow\pi^0+\text{e}^++\text{e}^-$ 
via $\eta\rightarrow\pi^0+\gamma^*$ results in 
\begin{align}
\label{eq:brvirtual}
	\frac{\Gamma\left(\eta\rightarrow\pi^0+\text{e}^++\text{e}^-\right)_\text{virtual}}
	{\Gamma\left(\eta\rightarrow\text{all}\right)}
	&< 7.5\times10^{-6} \nonumber \\
	&\qquad (\text{CL} = 90\,\%).
\end{align}
For comparison the assumption of a pure three-particle phase space distribution
of the ejectiles would lead to 
\begin{align}
\label{eq:brphase}
	\frac{\Gamma\left(\eta\rightarrow\pi^0+\text{e}^++\text{e}^-\right)_\text{phase}}
	{\Gamma\left(\eta\rightarrow\text{all}\right)}
	&< 9.5\times10^{-6} \nonumber \\
	&\qquad (\text{CL} = 90\,\%).
\end{align}
\par
These values are smaller than the previous upper limit of
$4.5\times10^{-5}$ (CL = $90\,\%$) \cite{Jane1975} by a factor of six and
five, respectively.

\section{Summary}
We have presented new studies with the WASA-at-COSY experiment on the $C$ parity violating 
$\eta$ meson decay $\eta\rightarrow\pi^0+\text{e}^++\text{e}^-$.
The obtained upper limit for the branching ratio of the decay
$\eta\rightarrow\pi^0+\text{e}^++\text{e}^-$ is smaller than
the previously available upper limit by a factor of five to six \cite{Jane1975}.
The results of the analysis are consistent with no events seen in data, and thus
give no hint on a $C$ violation in an electromagnetic
process. Similarly, no processes from physics beyond the Standard Model are
required to explain the results. 
\par
In order to further decrease this value and to continue the search for a $C$ parity 
violation in an electromagnetic process, additional data were collected
with WASA-at-COSY utilizing the production reaction
$\text{p}+\text{p}\rightarrow\text{p}+\text{p}+\eta$. Over three periods in
2008, 2010 and 2012 in total about $5\times10^8$ such events were recorded and
are currently being analyzed with regard to the decay
$\eta\rightarrow\pi^0+\text{e}^++\text{e}^-$.
\par
Besides a decay via one virtual photon, the decay
$\eta\rightarrow\pi^0+\text{e}^++\text{e}^-$ could possibly occur via
a hypothetical $C$ violating dark boson U with $m_\text{U} < 413\,\text{MeV}/c^2$  where the pertinent form factor is even further suppressed (i.e. the second term in its Taylor expansion vanishes) compared with the single-photon mechanism \cite{Kupsc2011}.
Investigations with regard to this decay process are currently ongoing for the
presented $\text{p}+\text{d}\rightarrow{^3\text{He}}+\eta$ data sets and the
$\text{p}+\text{p}\rightarrow\text{p}+\text{p}+\eta$ data sets recorded with
WASA-at-COSY.

\section*{Acknowledgements}
This work was supported in part by the EU Integrated Infrastructure Initiative
HadronPhysics Project under contract number RII3-CT-2004-506078; by the European
Commission under the 7th Framework Programme through the Research Infrastructures
action of the Capacities Programme, Call: FP7-INFRASTRUCTURES-2008-1, Grant
Agreement N.~227431; by the Polish National Science Centre through the grants
2016/23/B/ST2/00784, and the Foundation for Polish Science (MPD),
co-financed by the European Union within the European Regional Development Fund.
We gratefully acknowledge the support given by the Swedish Research Council,
the Knut and Alice Wallenberg Foundation, and the For\-schungs\-zen\-trum J\"u\-lich FFE
Funding Program. This work is based on the PhD thesis of Florian Sebastian
Bergmann.

Finally we thank all former WASA-at-COSY collaboration members for their
contribution to the success of the measurements, as well as the crew of the
COSY accelerator for their support during both measurement periods.

\section*{References}

\bibliography{mybibfile}

\end{document}